\documentclass[12pt]{article}    % onecolumn (standardformat)

\usepackage[margin=0.75in]{geometry} %In Folder
\usepackage{amsmath,amssymb,amsthm}
\newtheorem{theorem}{Theorem}[section]

\theoremstyle{remark}

\newcommand{\be}{\begin{equation}}
\newcommand{\ee}{\end{equation}}

\newcommand{\bea}{\begin{eqnarray}}
\newcommand{\eea}{\end{eqnarray}}

\numberwithin{equation}{section}
\begin{document}

\title{On the variance of linear statistics of Hermitian random matrices}
\author{Chao Min and Yang Chen\\
(chaominrunner@gmail.com, yangbrookchen@yahoo.co.uk)\\
Department of Mathematics, University of Macau,\\
Avenida da Universidade, Taipa, Macau, China}


\date{\today}
\maketitle
\begin{abstract}
Linear statistics, a random variable build out of the sum of the evaluation of functions at the eigenvalues of a $N\times N$ random matrix, $\sum_{j=1}^{N}f(x_{j})$ or $\mathrm{tr} f(M)$, is an ubiquitous statistical characteristics in random matrix theory.

Hermitian random matrix ensembles, under the eigenvalue-eigenvector decompositions give rise to the joint probability density functions of $N$ random variables.

We show that if $f(\cdot)$ is a polynomial of degree $K$, then the variance of $\mathrm{tr} f(M)$, is of the form, $\sum_{n=1}^{K}n(d_{n})^{2}$, and $d_{n}$ is related to the expansion coefficients $c_{n}$ of the polynomial
$f(x)=\sum_{n=0}^{K}c_{n}\widehat{P}_{n}(x)$, where $\widehat{P}_{n}(x)$ are polynomials of degree $n$, orthogonal with respect to the weights $\frac{1}{\sqrt{(b-x)(x-a)}}$, $\sqrt{(b-x)(x-a)}$, $\frac{\sqrt{(b-x)(x-a)}}{x},\; (0<a<x<b)$, $\frac{\sqrt{(b-x)(x-a)}}{x(1-x)},\\ (0<a<x<b<1)$, respectively.
\end{abstract}

\section{Introduction}
In the application of the theory of random matrices, we often encounter the random variable
$$
Q:=\text{tr\thinspace }f(M),
$$
where $f(M)$ is a real-valued function of the $N\times N$ random matrix $M$. In this paper, we will suppose $f(\cdot)$ to be a polynomial of degree $K$.

The space of matrices has the probability measure \cite{Mehta}
$$
\mbox{Prob}(M)\mbox{d}M:=\exp[-\text{tr\thinspace }v(M)]\mbox{d}M=C_{N}^{(\beta)}\prod_{1\leq j<k\leq N}|x_j-x_k|^\beta\prod_{1\leq l\leq N}e^{-v(x_{l})}\mbox{d}x_{l}.
$$
Here $\{x_{j}:1\leq j\leq N\}$ are the eigenvalues, $\beta=1,2,4$ are for matrices with orthogonal, unitary and symplectic symmetries respectively, $v(\cdot)$ is the potential
and $C_{N}^{(\beta)}$ is the normalization constant. In this paper we shall
only deal with the Hermitian case, i.e. $\beta=2$. For the purpose of this
paper we shall assume that $v(x)$ is convex, and therefore $v''(x)$ is positive on a set of positive measure.

In the limit of large $N$, the collection of eigenvalues can be approximated as a continuous fluid with a density supported in a single interval $(a, b)$.
We find the variance of $Q$ is \cite{Chen}
\be
\mathcal {V}=\frac{1}{2\pi^{2}}\int_{a}^{b}dx\frac{f(x)}{\sqrt{(b-x)(x-a)}}P\int_{a}^{b}dy\frac{\sqrt{(b-y)(y-a)}}{x-y}f'(y), \label{va}
\ee
where $P$ represents the principal value integral.

In this paper, we will consider four kinds of weight functions, all supported on [a,b], and write $f(\cdot)$ as the linear combination of the corresponding orthogonal polynomials, and obtain the relation between the variance and the coefficients, $c_{n}$. These weights are motivated by the equilibrium densities of the Jacobi ensembles (with parameters $\alpha=\beta=0$), Gaussian ensembles, Laguerre ensembles and the Jacobi ensembles (with general parameters), respectively.

In the paper \cite{Cabanal}, it was shown that if $f(x)=\sum_{n=0}^{K}c_{n}T_{n}(x)$, and $c_{n}$ the expansion coefficients of $f(\cdot)$ in terms of the Chebyshev polynomials of the first kind, then an equivalent of (\ref{va}) holds for\\
$a=-2, b=2$. Turning the table around, we show that starting from $f(x)=\sum_{n=0}^{K}c_{n}\widehat{P}_{n}(x)$, where $\widehat{P}_{n}(x)$ are polynomials of degree $n$, orthogonal with respect to the weights $\frac{1}{\sqrt{(b-x)(x-a)}}$, $\sqrt{(b-x)(x-a)}$, $\frac{\sqrt{(b-x)(x-a)}}{x},\;(0<a<x<b)$, $\frac{\sqrt{(b-x)(x-a)}}{x(1-x)},\;(0<a<x<b<1)$, respectively, then the variance from (\ref{va}) is the quadratic form,
$\sum_{m=1}^{K}\sum_{n=1}^{K}c_{m}c_{n}R(m,n)$, and when diagonalized becomes $\sum_{n=1}^{K}n(d_{n})^{2}$.

We would like to point out, the first two kinds of weights above correspond to the translated Chebyshev polynomials of the first kind and the second kind respectively. The last two kinds of weights correspond to the orthogonal polynomials which can be represented as the combination of the translated Chebyshev polynomials. Moreover, we want to say, the last two kinds of weights play an important role in the information theory of MIMO systems, which are just the eigenvalue densities of the single-user MIMO mutual information and multiuser MIMO mutual information respectively \cite{Chen2012}.

Now we recall the Chebyshev polynomials, which are crucial for our discussion throughout this paper. The Chebyshev polynomials of the first kind are defined by the recurrence relation
$$
T_{n}(x)=2x\:T_{n-1}(x)-T_{n-2}(x),\;\;n=2,3,\ldots
$$
with
$$
T_{0}(x)=1,\quad T_{1}(x)=x,
$$
and satisfy the orthogonality condition
$$
\int_{-1}^{1}T_{m}(x)T_{n}(x)\frac{dx}{\sqrt{1-x^{2}}}=\left\{
\begin{aligned}
&0,&m\neq n;\\
&\frac{\pi}{2}, &m=n\neq0;\\
&\pi, &m=n=0.
\end{aligned}
\right.
$$

Similarly, the Chebyshev polynomials of the second kind are defined by the relation
$$
U_{n}(x)=2x\:U_{n-1}(x)-U_{n-2}(x),\;\;n=2, 3, \ldots
$$
with
$$
U_{0}(x)=1,\quad U_{1}(x)=2x,
$$
and satisfy the orthogonality condition
$$
\int_{-1}^{1}U_{m}(x)U_{n}(x)\sqrt{1-x^{2}}dx=\left\{
\begin{aligned}
&0,&m\neq n;\\
&\frac{\pi}{2}, &m=n.
\end{aligned}
\right.
$$
More information on Chebyshev polynomials can be found in \cite{Szego,Andrews,Wang,Lebedev,Beals,Chihara,Gradshteyn}.

Now we introduce the translated Chebyshev polynomials. Define
$$
\widehat{T}_{n}(x):=T_{n}\left(\frac{2}{b-a}x-\frac{b+a}{b-a}\right),
$$
$$
\widehat{U}_{n}(x):=U_{n}\left(\frac{2}{b-a}x-\frac{b+a}{b-a}\right),
$$
then we have the relation
$$
\widehat{T}_{n}(x)=2\left(\frac{2}{b-a}x-\frac{b+a}{b-a}\right)\widehat{T}_{n-1}(x)-\widehat{T}_{n-2}(x),\;\;n=2,3,\ldots
$$
with the first two terms
$$
\widehat{T}_{0}(x)=1,\quad \widehat{T}_{1}(x)=\frac{2}{b-a}x-\frac{b+a}{b-a},
$$
and
$$
\widehat{U}_{n}(x)=2\left(\frac{2}{b-a}x-\frac{b+a}{b-a}\right)\widehat{U}_{n-1}(x)-\widehat{U}_{n-2}(x),\;\;n=2,3,\ldots
$$
with the first two terms
$$
\widehat{U}_{0}(x)=1,\quad \widehat{U}_{1}(x)=2\left(\frac{2}{b-a}x-\frac{b+a}{b-a}\right).
$$
The orthogonality conditions can be written respectively as
$$
\int_{a}^{b}\widehat{T}_{m}(x)\widehat{T}_{n}(x)\frac{dx}{\sqrt{(b-x)(x-a)}}=\left\{
\begin{aligned}
&0,&m\neq n;\\
&\frac{\pi}{2}, &m=n\neq0;\\
&\pi, &m=n=0,
\end{aligned}
\right.
$$
and
$$
\int_{a}^{b}\widehat{U}_{m}(x)\widehat{U}_{n}(x)\sqrt{(b-x)(x-a)}dx=\left\{
\begin{aligned}
&0,&m\neq n;\\
&\frac{\pi}{8}(b-a)^{2}, &m=n.
\end{aligned}
\right.
$$

\section{On the variance of linear statistics}
\subsection{On the weight $\frac{1}{\sqrt{(b-x)(x-a)}}$}
Now we suppose $f(x)$ is a polynomial of degree $K$, and write it as the linear combination of the translated Chebyshev polynomials $\widehat{T}_{n}(x)$, i.e.,
$$
f(x)=\sum_{n=0}^{K}c_{n}\widehat{T}_{n}(x).
$$
Since $T_{n}'(x)=n\:U_{n-1}(x),\;\;n=1,2,\ldots,$ see \cite{Gradshteyn} (page 995, 8.949(1)), we find
$$
f'(x)=\frac{2}{b-a}\sum_{n=1}^{K}n c_{n}\widehat{U}_{n-1}(x).
$$
It follows from (\ref{va}) that,
\be
\mathcal {V}=\frac{1}{\pi^{2}(b-a)}\int_{a}^{b}dx\frac{\sum_{m=0}^{K}c_{m}\widehat{T}_{m}(x)}{\sqrt{(b-x)(x-a)}}\sum_{n=1}^{K}nc_{n}P\int_{a}^{b}\frac{\sqrt{(b-y)(y-a)}}{x-y}
\widehat{U}_{n-1}(y)dy. \label{for}
\ee
Let
$$
x=\frac{b-a}{2}\tau+\frac{b+a}{2},\;\;y=\frac{b-a}{2}t+\frac{b+a}{2},
$$
then (\ref{for}) becomes,
\bea
\mathcal {V}
&=&\frac{1}{2\pi^{2}}\int_{-1}^{1}d\tau\frac{\sum_{m=0}^{K}c_{m}T_{m}(\tau)}{\sqrt{1-\tau^{2}}}\sum_{n=1}^{K}nc_{n}
P\int_{-1}^{1}\frac{\sqrt{1-t^{2}}}{\tau-t}U_{n-1}(t)dt\nonumber\\
&=&\frac{1}{2\pi}\sum_{m=0}^{K}\sum_{n=1}^{K}nc_{m}c_{n}\int_{-1}^{1}\frac{T_{m}(\tau)T_{n}(\tau)}{\sqrt{1-\tau^{2}}}d\tau\nonumber\\
&=&\frac{1}{4}\sum_{n=1}^{K}n(c_{n})^{2},\label{va1}
\eea
where we have used the formula \cite{Johansson}
\be
P\int_{-1}^{1}\frac{\sqrt{1-t^{2}}}{\tau-t}U_{n-1}(t)dt=\pi\:T_{n}(\tau),\;\;-1<\tau<1,\;\;n=1,2,\ldots. \label{utot}
\ee
in the first step.

$\mathbf{Remark}.$ The result (\ref{va1}) coincides with \cite{Cabanal} for the special case $a=-2, b=2$.

\subsection{On the weight $\sqrt{(b-x)(x-a)}$}
Now we consider another case, and represent $f(x)$ as the linear combination of the translated Chebyshev polynomials $\widehat{U}_{n}(x)$, i.e.,
$$
f(x)=\sum_{n=0}^{K}c_{n}\widehat{U}_{n}(x).
$$
Since $U_{n}'(x)=\frac{1}{1-x^{2}}\left[(n+1)U_{n-1}(x)-nx\:U_{n}(x)\right],\;\;n=1,2,\ldots,$ see \cite{Gradshteyn} (page 995, 8.949(6)), we obtain
$$
f'(x)=\frac{b-a}{2}\sum_{n=1}^{K}\frac{c_{n}}{(b-x)(x-a)}\left[(n+1)\widehat{U}_{n-1}(x)-n\left(\frac{2}{b-a}x-\frac{b+a}{b-a}\right)\widehat{U}_{n}(x)\right].
$$
It follows from (\ref{va}) that,
\be
\mathcal {V}=\frac{b-a}{4\pi^{2}}\sum_{m=0}^{K}\sum_{n=1}^{K}c_{m}c_{n}\int_{a}^{b}\frac{dx\:\widehat{U}_{m}(x)}{\sqrt{(b-x)(x-a)}}
P\int_{a}^{b}\frac{(n+1)\widehat{U}_{n-1}(y)-n\left(\frac{2}{b-a}y-\frac{b+a}{b-a}\right)\widehat{U}_{n}(y)}{\sqrt{(b-y)(y-a)}(x-y)}dy. \label{for1}
\ee
Let
$$
x=\frac{b-a}{2}\tau+\frac{b+a}{2},\;\;y=\frac{b-a}{2}t+\frac{b+a}{2},
$$
then (\ref{for1}) becomes,
\be
\mathcal {V}=\frac{1}{2\pi^{2}}\sum_{m=0}^{K}\sum_{n=1}^{K}c_{m}c_{n}\int_{-1}^{1}\frac{d\tau\:U_{m}(\tau)}{\sqrt{1-\tau^{2}}}
P\int_{-1}^{1}\frac{(n+1)U_{n-1}(t)-nt\:U_{n}(t)}{\sqrt{1-t^{2}}(\tau-t)}dt. \label{var}
\ee
To proceed further, we need to calculate two integrals, the first being
$$
P\int_{-1}^{1}\frac{U_{n-1}(t)dt}{\sqrt{1-t^{2}}(\tau-t)}
$$
and the second,
$$
P\int_{-1}^{1}\frac{t\:U_{n}(t)dt}{\sqrt{1-t^{2}}(\tau-t)}.
$$
We see
\bea
&&P\int_{-1}^{1}\frac{U_{n-1}(t)dt}{\sqrt{1-t^{2}}(\tau-t)}
=\frac{1}{2}P\int_{-1}^{1}\frac{\sqrt{1-t^{2}}\:U_{n-1}(t)}{\tau-t}\left(\frac{1}{1-t}+\frac{1}{1+t}\right)dt
\nonumber\\
&=&\frac{1}{2}\left[\frac{2}{1-\tau^{2}}P\int_{-1}^{1}\frac{\sqrt{1-t^{2}}\:U_{n-1}(t)}{\tau-t}dt
-\frac{1}{1-\tau}\int_{-1}^{1}\sqrt{\frac{1+t}{1-t}}\:U_{n-1}(t)dt+\frac{1}{1+\tau}\int_{-1}^{1}\sqrt{\frac{1-t}{1+t}}\:U_{n-1}(t)dt\right]\nonumber\\
&=&\frac{\pi\: T_{n}(\tau)}{1-\tau^{2}}-\frac{\pi}{2(1-\tau)}-\frac{(-1)^{n}\pi}{2(1+\tau)}, \label{prin}
\eea
where we have used (\ref{utot}) and
\be
\int_{-1}^{1}\sqrt{\frac{1+t}{1-t}}\:U_{n-1}(t)dt=\pi,\;\;n=1,2,\ldots, \label{eq4}
\ee
\be
\int_{-1}^{1}\sqrt{\frac{1-t}{1+t}}\:U_{n-1}(t)dt=(-1)^{n-1}\pi,\;\;n=1,2,\ldots. \label{eq5}
\ee
Next, we compute,
\bea
&&P\int_{-1}^{1}\frac{t\:U_{n}(t)dt}{\sqrt{1-t^{2}}(\tau-t)}\nonumber\\
&=&\tau\:P\int_{-1}^{1}\frac{U_{n}(t)dt}{\sqrt{1-t^{2}}(\tau-t)}
-\int_{-1}^{1}\frac{U_{n}(t)dt}{\sqrt{1-t^{2}}}\nonumber\\
&=&\frac{\pi\:\tau\: T_{n+1}(\tau)}{1-\tau^{2}}-\frac{\pi}{2(1-\tau)}-\frac{(-1)^{n}\pi}{2(1+\tau)}, \label{next}
\eea
where use has been made of (\ref{prin}) and
\be
\int_{-1}^{1}\frac{U_{n}(t)dt}{\sqrt{1-t^{2}}}=\frac{1+(-1)^{n}}{2}\pi,\;\;n=0,1,2,\ldots. \label{eq6}
\ee

$\mathbf{Remark}.$ The above integrals involving $U_{n}(t)$, (\ref{eq4}), (\ref{eq5}) and (\ref{eq6}), can be easily obtained with the substituting $t=\cos\theta,\; 0\leq\theta\leq \pi$.

It follows from (\ref{prin}) and (\ref{next}) that
$$
P\int_{-1}^{1}\frac{(n+1)U_{n-1}(t)-nt\:U_{n}(t)}{\sqrt{1-t^{2}}(\tau-t)}dt=\frac{\pi}{1-\tau^{2}}\left[(n+1)T_{n}(\tau)-n\tau\:T_{n+1}(\tau)-\frac{1+\tau}{2}
-\frac{(-1)^{n}(1-\tau)}{2}\right].
$$
Hence by (\ref{var}),
$$
\mathcal {V}=\frac{1}{4}\sum_{m=1}^{K}\sum_{n=1}^{K}c_{m}c_{n}R(m,n),
$$
where
\bea
R(m,n)&=&\frac{2}{\pi}\int_{-1}^{1}\frac{U_{m}(\tau)}{(1-\tau^{2})^{\frac{3}{2}}}\left[(n+1)T_{n}(\tau)-n\tau\:T_{n+1}(\tau)-\frac{1+\tau}{2}
-\frac{(-1)^{n}(1-\tau)}{2}\right]d\tau\nonumber\\
&=&
\left\{
\begin{aligned}
&0, &m=2,4,6,\ldots,\; n=1,3,5,\ldots;\\
&n^{2}+2n, &m=2,4,6,\ldots,\; n=2,4,\ldots,m;\\
&m^{2}+2m, &\qquad m=2,4,6,\ldots,\; n=m+2,m+4,\ldots;\\
&0, &m=1,3,5,\ldots,\; n=2,4,6,\ldots;\\
&(n+1)^{2}, &m=1,3,5,\ldots,\; n=1,3,\ldots,m;\\
&(m+1)^{2}, &m=1,3,5,\ldots,\; n=m+2,m+4,\ldots.
\end{aligned}
\right.\nonumber
\eea
Note that $R(m,n)$ is symmetric in $m$ and $n$. To bring the above quadratic form to a diagonal form, we use the rotation,
$$
\begin{pmatrix}
c_{1}\\
c_{2}\\
c_{3}\\
c_{4}\\
\vdots\\
c_{K-2}\\
c_{K-1}\\
c_{K}
\end{pmatrix}
=\begin{pmatrix}
1&0&-1&0&\cdots&0&0&0\\
0&1&0&-1&\cdots&0&0&0\\
0&0&1&0&\cdots&0&0&0\\
0&0&0&1&\cdots&0&0&0\\
\vdots&\vdots&\vdots&\vdots&&\vdots&\vdots&\vdots\\
0&0&0&0&\cdots&1&0&-1\\
0&0&0&0&\cdots&0&1&0\\
0&0&0&0&\cdots&0&0&1
\end{pmatrix}
\begin{pmatrix}
d_{1}\\
d_{2}\\
d_{3}\\
d_{4}\\
\vdots\\
d_{K-2}\\
d_{K-1}\\
d_{K}
\end{pmatrix},
$$
and obtain
$$
\mathcal {V}=\sum_{n=1}^{K}n(d_{n})^{2}.
$$

\subsection{On the weight $\frac{\sqrt{(b-x)(x-a)}}{x}$}
Now we consider the weight function
$$
w(x)=\frac{\sqrt{(b-x)(x-a)}}{x},\;\;0<a<x<b.
$$
Let
$$
x=\frac{b-a}{2}\tau+\frac{b+a}{2},\;\;-1<\tau<1,
$$
then we have
$$
w(x)=\frac{\sqrt{1-\tau^{2}}}{\rho(\tau)},
$$
where
\be
\rho(\tau)=\tau+\frac{b+a}{b-a}>0. \label{rho}
\ee
We introduce here a theorem of $\mathrm{Szeg\ddot{o}}$ \cite{Szego}, which is essential to construct the orthogonal polynomials we need in this and the next sections.
\begin{theorem}
Let $\rho(\tau)$ be a polynomial of degree $l$ and positive in $[-1,1]$. Let $\rho(\cos\theta)=|h(e^{i\theta})|^{2}$, where $h(z)$ is a polynomial of degree $l$ with real coefficients, and $h(z)\neq 0$ in $|z|<1$, $h(0)>0$. Writing $h(e^{i\theta})=c(\theta)+is(\theta)$, $c(\theta)$ and $s(\theta)$ real, we have the orthonormal polynomials with respect to the weight function $\frac{\sqrt{1-\tau^{2}}}{\rho(\tau)}$,
$$
P_{n}(\cos\theta)=\sqrt{\frac{2}{\pi}}\left[c(\theta)\frac{\sin(n+1)\theta}{\sin\theta}-s(\theta)\frac{\cos(n+1)\theta}{\sin\theta}\right],\;\;n>\frac{l}{2}-1. \label{thm}
$$
\end{theorem}

From (\ref{rho}),
$$
\rho(\cos\theta)=\cos\theta+\frac{b+a}{b-a}.
$$
Write $\rho(\cos\theta)$ in the form
$$
\rho(\cos\theta)=|h(e^{i\theta})|^{2},
$$
where
\be
h(e^{i\theta})=A e^{i\theta}+B, \label{h}
\ee
such that $h(z)\neq0$ in $|z|<1$, $A$ is real and $B>0$.

With the above equalities, we have the equation
$$
(A e^{i\theta}+B)\overline{(A e^{i\theta}+B)}=\cos\theta+\frac{b+a}{b-a},
$$
which is equivalent to the equations
$$
\left\{\begin{aligned}
&2AB=1\\
&A^{2}+B^{2}=\frac{b+a}{b-a}.
\end{aligned}
\right.
$$
A simple computation shows that
$$
\left\{\begin{aligned}
&A=\frac{\sqrt{b}-\sqrt{a}}{\sqrt{2(b-a)}}\\
&B=\frac{\sqrt{b}+\sqrt{a}}{\sqrt{2(b-a)}}.
\end{aligned}
\right.
$$
It follows from (\ref{h}) that
$$
h(e^{i\theta})=c(\theta)+is(\theta),
$$
where
$$
c(\theta)=A\cos\theta+B,
$$
$$
s(\theta)=A\sin\theta.
$$

According to Theorem \ref{thm}, we obtain the orthonormal polynomials with respect to
the weight function $\frac{\sqrt{1-\tau^{2}}}{\rho(\tau)}$,
$$
P_{n}(\cos\theta)=\sqrt{\frac{2}{\pi}}\left[(A\cos\theta+B)\frac{\sin(n+1)\theta}{\sin\theta}-A\cos(n+1)\theta\right],\;\;n=0,1,2,\ldots.
$$
Substituting $\tau$ for $\cos\theta$, we have
\bea
P_{n}(\tau)&=&\sqrt{\frac{2}{\pi}}\left[(A\tau+B)U_{n}(\tau)-A\: T_{n+1}(\tau)\right]\nonumber\\
&=&\frac{\left[\left(\sqrt{b}-\sqrt{a}\right)\tau+\sqrt{b}+\sqrt{a}\right]U_{n}(\tau)-\left(\sqrt{b}-\sqrt{a}\right)T_{n+1}(\tau)}{\sqrt{\pi(b-a)}}.\nonumber
\eea
It is easy to see that,
$$
P_{n}'(\tau)=\frac{(n+1)\left[\left(\sqrt{b}-\sqrt{a}\right)\tau+\sqrt{b}+\sqrt{a}\right]U_{n-1}(\tau)
-n\left[\left(\sqrt{b}+\sqrt{a}\right)\tau+\sqrt{b}-\sqrt{a}\right]U_{n}(\tau)}{\sqrt{\pi(b-a)}(1-\tau^{2})}.
$$
Let
\bea
\widehat{P}_{n}(x):&=&\sqrt{\pi(b-a)}P_{n}\left(\frac{2}{b-a}x-\frac{b+a}{b-a}\right)\nonumber\\
&=&\frac{2\left(x+\sqrt{ab}\right)}{\sqrt{b}+\sqrt{a}}\widehat{U}_{n}(x)
-\left(\sqrt{b}-\sqrt{a}\right)\widehat{T}_{n+1}(x),\nonumber
\eea
then we have the orthogonality condition
$$
\int_{a}^{b}\widehat{P}_{m}(x)\widehat{P}_{n}(x)\frac{\sqrt{(b-x)(x-a)}}{x}dx=\frac{\pi(b-a)^{2}}{2}\delta_{mn},
$$
since
$$
\int_{-1}^{1}P_{m}(\tau)P_{n}(\tau)\frac{\sqrt{1-\tau^{2}}}{\tau+\frac{b+a}{b-a}}d\tau=\delta_{mn}.
$$

Now let $f(x)$ be the linear combination of the orthogonal polynomials $\widehat{P}_{n}(x)$, i.e.,
$$
f(x)=\sum_{n=0}^{K}c_{n}\widehat{P}_{n}(x),
$$
then we find
$$
f'(x)=\sum_{n=1}^{K}c_{n}\widehat{P}_{n}'(x)=\frac{2\sqrt{\pi}}{\sqrt{b-a}}\sum_{n=1}^{K}c_{n}P_{n}'\left(\frac{2}{b-a}x-\frac{b+a}{b-a}\right).
$$
Let
$$
x=\frac{b-a}{2}\tau+\frac{b+a}{2},\;\;y=\frac{b-a}{2}t+\frac{b+a}{2},
$$
then it follows from (\ref{va}) that
\bea
\mathcal {V}&=&\frac{b-a}{2\pi}\int_{-1}^{1}d\tau\frac{\sum_{m=0}^{K}c_{m}P_{m}(\tau)}{\sqrt{1-\tau^{2}}}P\int_{-1}^{1}\frac{\sqrt{1-t^{2}}}{\tau-t}
\sum_{n=1}^{K}c_{n}P_{n}'(t)dt\nonumber\\
&=&\frac{\sqrt{\pi(b-a)}}{2\pi^{2}}\sum_{m=0}^{K}\sum_{n=1}^{K}c_{m}c_{n}\int_{-1}^{1}\frac{P_{m}(\tau)g_{n}(\tau)}{\sqrt{1-\tau^{2}}}d\tau,\nonumber
\eea
where
\bea
g_{n}(\tau):&=&P\int_{-1}^{1}\frac{1}{\sqrt{1-t^{2}}(\tau-t)}
\bigg\{(n+1)\left[\left(\sqrt{b}-\sqrt{a}\right)t+\sqrt{b}+\sqrt{a}\right]U_{n-1}(t)\nonumber\\
&-&n\left[\left(\sqrt{b}+\sqrt{a}\right)t+\sqrt{b}-\sqrt{a}\right]U_{n}(t)\bigg\}dt.\nonumber
\eea
By using (\ref{prin}) and (\ref{next}), we readily obtain
\bea
g_{n}(\tau)
&=&\frac{\pi}{1-\tau^{2}}\bigg\{(n+1)\left[\left(\sqrt{b}-\sqrt{a}\right)\tau+\sqrt{b}+\sqrt{a}\right]T_{n}(\tau)
-n\left[\left(\sqrt{b}+\sqrt{a}\right)\tau+\sqrt{b}-\sqrt{a}\right]T_{n+1}(\tau)\nonumber\\
&-&\sqrt{b}(1+\tau)-(-1)^{n}\sqrt{a}(1-\tau)\bigg\}.\nonumber
\eea
It follows that
$$
\mathcal {V}=\frac{1}{2}\sum_{m=1}^{K}\sum_{n=1}^{K}c_{m}c_{n}R(m,n),
$$
where if $m=2,4,6,\ldots$,
$$
R(m,n)=\left\{
\begin{aligned}
&(b+a)n^{2}+\left(\sqrt{b}+\sqrt{a}\right)^{2}n, &n=2,4,\ldots,m;\\
&(b+a)m^{2}+\left(\sqrt{b}+\sqrt{a}\right)^{2}m,\qquad &n=m+2,m+4,\ldots;\\
&(b-a)\left(n^{2}+n\right), &n=1,3,\ldots,m-1;\\
&(b-a)\left(m^{2}+m\right), &n=m+1,m+3,\ldots,\\
\end{aligned}
\right.
$$
if $m=1,3,5,\ldots$,
$$
R(m,n)=\left\{
\begin{aligned}
&(b-a)\left(n^{2}+n\right), &n=2,4,\ldots,m-1;\\
&(b-a)\left(m^{2}+m\right), &n=m+1,m+3,\ldots;\\
&(b+a)n^{2}+\left(\sqrt{b}+\sqrt{a}\right)^{2}n+2\sqrt{ab}, &n=1,3,\ldots,m;\\
&(b+a)m^{2}+\left(\sqrt{b}+\sqrt{a}\right)^{2}m+2\sqrt{ab},\qquad &n=m+2,m+4,\ldots.\\
\end{aligned}
\right.
$$
Note that $R(m,n)$ is symmetric in $m$ and $n$. Let
$$
\begin{pmatrix}
c_{1}\\[5pt]
c_{2}\\[5pt]
c_{3}\\[5pt]
c_{4}\\[5pt]
\vdots\\
c_{K}
\end{pmatrix}
=\begin{pmatrix}
\frac{1}{\sqrt{b}+\sqrt{a}}&-\frac{\sqrt{b}-\sqrt{a}}{\left(\sqrt{b}+\sqrt{a}\right)^{2}}&-\frac{4\sqrt{ab}}{\left(\sqrt{b}+\sqrt{a}\right)^{3}}
&\frac{4\sqrt{ab}\left(\sqrt{b}-\sqrt{a}\right)}{\left(\sqrt{b}+\sqrt{a}\right)^{4}}&\cdots&(-1)^{K}\frac{4\sqrt{ab}\left(\sqrt{b}-\sqrt{a}\right)^{K-3}}
{\left(\sqrt{b}+\sqrt{a}\right)^{K}}\\
0&\frac{1}{\sqrt{b}+\sqrt{a}}&-\frac{\sqrt{b}-\sqrt{a}}{\left(\sqrt{b}+\sqrt{a}\right)^{2}}&-\frac{4\sqrt{ab}}{\left(\sqrt{b}+\sqrt{a}\right)^{3}}
&\cdots&(-1)^{K-1}\frac{4\sqrt{ab}\left(\sqrt{b}-\sqrt{a}\right)^{K-4}}{\left(\sqrt{b}+\sqrt{a}\right)^{K-1}}\\
0&0&\frac{1}{\sqrt{b}+\sqrt{a}}&-\frac{\sqrt{b}-\sqrt{a}}{\left(\sqrt{b}+\sqrt{a}\right)^{2}}
&\cdots&(-1)^{K-2}\frac{4\sqrt{ab}\left(\sqrt{b}-\sqrt{a}\right)^{K-5}}{\left(\sqrt{b}+\sqrt{a}\right)^{K-2}}\\
0&0&0&\frac{1}{\sqrt{b}+\sqrt{a}}&\cdots&(-1)^{K-3}\frac{4\sqrt{ab}\left(\sqrt{b}-\sqrt{a}\right)^{K-6}}{\left(\sqrt{b}+\sqrt{a}\right)^{K-3}}\\
\vdots&\vdots&\vdots&\vdots&&\vdots\\
0&0&0&0&\cdots&\frac{1}{\sqrt{b}+\sqrt{a}}
\end{pmatrix}
\begin{pmatrix}
d_{1}\\[5pt]
d_{2}\\[5pt]
d_{3}\\[5pt]
d_{4}\\[5pt]
\vdots\\
d_{K}
\end{pmatrix},
$$
then we obtain
$$
\mathcal {V}=\sum_{n=1}^{K}n(d_{n})^{2}.
$$

\subsection{On the weight $\frac{\sqrt{(b-x)(x-a)}}{x(1-x)}$}
Now we consider another weight function
$$
w(x)=\frac{\sqrt{(b-x)(x-a)}}{x(1-x)},\;\;0<a<x<b<1.
$$
Let
$$
x=\frac{b-a}{2}\tau+\frac{b+a}{2},\;\;-1<\tau<1,
$$
then we have
$$
w(x)=\frac{\sqrt{1-\tau^{2}}}{\eta(\tau)},
$$
where
$$
\eta(\tau)=\frac{a-b}{2}\tau^{2}+(1-a-b)\tau+\frac{(a+b)(a+b-2)}{2(a-b)}>0,
$$
and hence we have
\be
\eta(\cos\theta)=\frac{a-b}{2}\cos^{2}\theta+(1-a-b)\cos\theta+\frac{(a+b)(a+b-2)}{2(a-b)}. \label{eq1}
\ee
Write $\eta(\cos\theta)$ in the form
\be
\eta(\cos\theta)=\left|h\left(e^{i\theta}\right)\right|^{2}, \label{eq2}
\ee
where
\be
h\left(e^{i\theta}\right)=A e^{2i\theta}+B e^{i\theta}+C, \label{eq3}
\ee
such that $h(z)\neq0$ in $|z|<1$, $A$ and $B$ are real and $C>0$.

From (\ref{eq1}), (\ref{eq2}) and (\ref{eq3}), we have the equation
$$
\left(A e^{2i\theta}+B e^{i\theta}+C\right)\overline{\left(A e^{2i\theta}+B e^{i\theta}+C\right)}=\frac{a-b}{2}\cos^{2}\theta+(1-a-b)\cos\theta+\frac{(a+b)(a+b-2)}{2(a-b)},
$$
which is equivalent to the equations
$$
\left\{\begin{aligned}
&4AC=\frac{a-b}{2}\\
&2(A+C)B=1-a-b\\
&(A-C)^{2}+B^{2}=\frac{(a+b)(a+b-2)}{2(a-b)}.
\end{aligned}
\right.
$$
An elaborate computation shows that
$$
\left\{\begin{aligned}
&A=\frac{\left(\sqrt{b}-\sqrt{a}\right)\left(\sqrt{1-b}-\sqrt{1-a}\right)}{2\sqrt{2(b-a)}}\\[8pt]
&B=\frac{\sqrt{b(1-b)}-\sqrt{a(1-a)}}{\sqrt{2(b-a)}}\\[8pt]
&C=\frac{\left(\sqrt{b}+\sqrt{a}\right)\left(\sqrt{1-b}+\sqrt{1-a}\right)}{2\sqrt{2(b-a)}}.
\end{aligned}
\right.
$$
Write $h(e^{i\theta})$ in the form
$$
h(e^{i\theta})=c(\theta)+is(\theta),
$$
where
$$
c(\theta)=A\cos2\theta+B\cos\theta+C,
$$
$$
s(\theta)=A\sin2\theta+B\sin\theta.
$$
According to Theorem \ref{thm}, we obtain the orthonormal polynomials with respect to the
weight function $\frac{\sqrt{1-\tau^{2}}}{\eta(\tau)}$, i.e.,
$$
P_{n}(\cos\theta)=\sqrt{\frac{2}{\pi}}\left\{c(\theta)\frac{\sin(n+1)\theta}{\sin\theta}-s(\theta)\frac{\cos(n+1)\theta}{\sin\theta}\right\},\;\;n=1,2,\ldots.
$$
Substituting $\tau$ for $\cos\theta$ and we have
\bea
P_{n}(\tau)
&=&\sqrt{\frac{2}{\pi}}\left[\left(2A\tau^{2}+B\tau+C-A\right)U_{n}(\tau)-(2A\tau+B)T_{n+1}(\tau)\right]\nonumber\\
&=&\sqrt{\frac{2}{\pi}}\left[A\:U_{n-2}(\tau)+B\:U_{n-1}(\tau)+C\:U_{n}(\tau)\right]\nonumber\\
&=&\frac{1}{2\sqrt{\pi(b-a)}}\bigg\{\left(\sqrt{b}-\sqrt{a}\right)\left(\sqrt{1-b}-\sqrt{1-a}\right)U_{n-2}(\tau)+2\left[\sqrt{b(1-b)}-\sqrt{a(1-a)}\right]U_{n-1}(\tau)\nonumber\\
&+&\left(\sqrt{b}+\sqrt{a}\right)\left(\sqrt{1-b}+\sqrt{1-a}\right)U_{n}(\tau)\bigg\}.\nonumber
\eea
It follows that
\bea
P_{n}'(\tau)&=&\frac{1}{2\sqrt{\pi(b-a)}(1-\tau^{2})}\bigg\{\left(\sqrt{b}-\sqrt{a}\right)\left(\sqrt{1-b}-\sqrt{1-a}\right)\Big[(n-1)U_{n-3}(\tau)-(n-2)\tau U_{n-2}(\tau)\Big]\nonumber\\
&+&2\left[\sqrt{b(1-b)}-\sqrt{a(1-a)}\right]\Big[nU_{n-2}(\tau)-(n-1)\tau U_{n-1}(\tau)\Big]\nonumber\\
&+&\left(\sqrt{b}+\sqrt{a}\right)\left(\sqrt{1-b}+\sqrt{1-a}\right)\Big[(n+1)U_{n-1}(\tau)-n\tau U_{n}(\tau)\Big]\bigg\}.\nonumber
\eea
Let
\bea
\widehat{P}_{n}(x):&=&2\sqrt{\pi(b-a)}P_{n}\left(\frac{2}{b-a}x-\frac{b+a}{b-a}\right)\nonumber\\
&=&\left(\sqrt{b}-\sqrt{a}\right)\left(\sqrt{1-b}-\sqrt{1-a}\right)\widehat{U}_{n-2}(x)+2\left[\sqrt{b(1-b)}-\sqrt{a(1-a)}\right]\widehat{U}_{n-1}(x)\nonumber\\
&+&\left(\sqrt{b}+\sqrt{a}\right)\left(\sqrt{1-b}+\sqrt{1-a}\right)\widehat{U}_{n}(x),\;\;n=0,1,2,\ldots,\nonumber
\eea
then we have the orthogonality condition
$$
\int_{a}^{b}\widehat{P}_{m}(x)\widehat{P}_{n}(x)\frac{\sqrt{(b-x)(x-a)}}{x(1-x)}dx=2\pi(b-a)^{2}\delta_{mn},\;\;m=0,1,2,\ldots,\; n=1,2,\ldots,
$$
since
$$
\int_{-1}^{1}P_{m}(\tau)P_{n}(\tau)\frac{\sqrt{1-\tau^{2}}}{\eta(\tau)}d\tau=\delta_{mn}.
$$

Let $f(x)$ be the linear combination of the orthogonal polynomials $\widehat{P}_{n}(x)$, i.e.,
$$
f(x)=\sum_{n=0}^{K}c_{n}\widehat{P}_{n}(x),
$$
then we have
$$
f'(x)=\sum_{n=1}^{K}c_{n}\widehat{P}_{n}'(x)=\frac{4\sqrt{\pi}}{\sqrt{b-a}}\sum_{n=1}^{K}c_{n}P_{n}'\left(\frac{2}{b-a}x-\frac{b+a}{b-a}\right).
$$
Let
$$
x=\frac{b-a}{2}\tau+\frac{b+a}{2},\;\;y=\frac{b-a}{2}t+\frac{b+a}{2},
$$
then it follows from (\ref{va}) that
\bea
\mathcal {V}&=&\frac{2(b-a)}{\pi}\int_{-1}^{1}d\tau\frac{\sum_{m=0}^{K}c_{m}P_{m}(\tau)}{\sqrt{1-\tau^{2}}}P\int_{-1}^{1}\frac{\sqrt{1-t^{2}}}{\tau-t}
\sum_{n=1}^{K}c_{n}P_{n}'(t)dt\nonumber\\
&=&\frac{\sqrt{\pi(b-a)}}{\pi^{2}}\sum_{m=0}^{K}\sum_{n=1}^{K}c_{m}c_{n}\int_{-1}^{1}\frac{P_{m}(\tau)g_{n}(\tau)}{\sqrt{1-\tau^{2}}}d\tau,\nonumber
\eea
where
\bea
g_{n}(\tau):&=&P\int_{-1}^{1}\frac{1}{\sqrt{1-t^{2}}(\tau-t)}\bigg\{\left(\sqrt{b}-\sqrt{a}\right)\left(\sqrt{1-b}-\sqrt{1-a}\right)\Big[(n-1)U_{n-3}(t)-(n-2)t U_{n-2}(t)\Big]\nonumber\\
&+&2\left[\sqrt{b(1-b)}-\sqrt{a(1-a)}\right]\Big[nU_{n-2}(t)-(n-1)t U_{n-1}(t)\Big]\nonumber\\
&+&\left(\sqrt{b}+\sqrt{a}\right)\left(\sqrt{1-b}+\sqrt{1-a}\right)\Big[(n+1)U_{n-1}(t)-nt U_{n}(t)\Big]\bigg\}dt.\nonumber
\eea
By using (\ref{prin}) and (\ref{next}), a simple computation shows that
\bea
g_{n}(\tau)&=&\frac{\pi}{1-\tau^{2}}\bigg\{\left(\sqrt{b}-\sqrt{a}\right)\left(\sqrt{1-b}-\sqrt{1-a}\right)\Big[(n-1)T_{n-2}(\tau)-(n-2)\tau T_{n-1}(\tau)\Big]\nonumber\\
&+&2\left[\sqrt{b(1-b)}-\sqrt{a(1-a)}\right]\Big[n T_{n-1}(\tau)-(n-1)\tau T_{n}(\tau)\Big]\nonumber\\
&+&\left(\sqrt{b}+\sqrt{a}\right)\left(\sqrt{1-b}+\sqrt{1-a}\right)\Big[(n+1)T_{n}(\tau)-n\tau T_{n+1}(\tau)\Big]\nonumber\\
&-&2\sqrt{b(1-b)}(1+\tau)-2(-1)^{n}\sqrt{a(1-a)}(1-\tau)\bigg\}.\nonumber
\eea
It follows that
$$
\mathcal {V}=\sum_{m=1}^{K}\sum_{n=1}^{K}c_{m}c_{n}R(m,n),
$$
where if $m=2,4,6,\ldots$,
$$
R(m,n)=\left\{
\begin{aligned}
&2n(b-a)\left[\sqrt{(1-a)(1-b)}-\sqrt{a b}-na-nb+n\right],\quad n=1,3,\ldots,m-1;\\[10pt]
&2n(a+b)\sqrt{(1-a)(1-b)}-2n\sqrt{a b}(a+b-2)-2n^{2}\left(a^{2}+b^{2}-a-b\right),\quad n=2,4,\ldots,m-2;\\[10pt]
&2n(a+b)\sqrt{(1-a)(1-b)}-2n\sqrt{a b}(a+b-2)-2n^{2}\left(a^{2}+b^{2}-a-b\right)+n(a-b)^{2},\quad n=m;\\[10pt]
&2m(b-a)\left[\sqrt{(1-a)(1-b)}-\sqrt{a b}-ma-mb+m\right],\quad n=m+1,m+3,\ldots;\\[10pt]
&2m(a+b)\sqrt{(1-a)(1-b)}-2m\sqrt{a b}(a+b-2)-2m^{2}\left(a^{2}+b^{2}-a-b\right), n=m+2,m+4,\ldots,
\end{aligned}
\right.
$$
if $m=1,3,5,\ldots$,
$$
R(m,n)=\left\{
\begin{aligned}
&2n(b-a)\left[\sqrt{(1-a)(1-b)}-\sqrt{a b}-na-nb+n\right], \quad n=2,4,\ldots,m-1;\\[10pt]
&2\sqrt{(1-a)(1-b)}\left(na+nb+2\sqrt{ab}\right)-2n\sqrt{ab}(a+b-2)-2n^{2}\left(a^{2}+b^{2}-a-b\right),\\
&n=1,3,\ldots,m-2;\\[10pt]
&2\sqrt{(1-a)(1-b)}\left(na+nb+2\sqrt{ab}\right)-2n\sqrt{ab}(a+b-2)-2n^{2}\left(a^{2}+b^{2}-a-b\right)\\
&+n(a-b)^{2},\quad n=m;\\[10pt]
&2m(b-a)\left[\sqrt{(1-a)(1-b)}-\sqrt{a b}-ma-mb+m\right],\quad n=m+1,m+3,\ldots;\\[10pt]
&2\sqrt{(1-a)(1-b)}\left(ma+mb+2\sqrt{ab}\right)-2m\sqrt{ab}(a+b-2)-2m^{2}\left(a^{2}+b^{2}-a-b\right),\\
&n=m+2,m+4,\ldots.
\end{aligned}
\right.
$$
Note that $R(m,n)$ is symmetric in $m$ and $n$, so we can also write $\mathcal {V}$ in the form as the sum of the square of new variables theoretically, but since the transformation is very complicated, we will only give an example here, $K=5$. Let
$$
\begin{pmatrix}
c_{1}\\
c_{2}\\
c_{3}\\
c_{4}\\
c_{5}
\end{pmatrix}
=\begin{pmatrix}
\alpha&\beta&\gamma&\xi&\zeta\\
0&\alpha&\beta&\gamma&\xi\\
0&0&\alpha&\beta&\gamma\\
0&0&0&\alpha&\beta\\
0&0&0&0&\alpha
\end{pmatrix}
\begin{pmatrix}
d_{1}\\
d_{2}\\
d_{3}\\
d_{4}\\
d_{5}
\end{pmatrix},
$$
where
$$
\alpha=\frac{1}{\left(\sqrt{a}+\sqrt{b}\right)\left(\sqrt{1-a}+\sqrt{1-b}\right)},
$$

$$
\beta=\frac{2\left[\sqrt{a(1-a)}-\sqrt{b(1-b)}\right]}
{\left(\sqrt{a}+\sqrt{b}\right)^{2}\left(\sqrt{1-a}+\sqrt{1-b}\right)^{2}},
$$

$$
\gamma=\frac{4\left[\sqrt{a(1-a)}-\sqrt{b(1-b)}\right]^{2}}{\left(\sqrt{a}+\sqrt{b}\right)^{3}\left(\sqrt{1-a}+\sqrt{1-b}\right)^{3}}
-\frac{2\left[\sqrt{a(1-a)}+\sqrt{b(1-b)}\right]}{\left(\sqrt{a}+\sqrt{b}\right)^{2}\left(\sqrt{1-a}+\sqrt{1-b}\right)^{2}},
$$

\bea
\xi&=&\frac{8\left[\sqrt{a(1-a)}-\sqrt{b(1-b)}\right]^{3}}{\left(\sqrt{a}+\sqrt{b}\right)^{4}\left(\sqrt{1-a}+\sqrt{1-b}\right)^{4}}
-\frac{4(a-b)(1-a-b)}{\left(\sqrt{a}+\sqrt{b}\right)^{3}\left(\sqrt{1-a}+\sqrt{1-b}\right)^{3}}\nonumber\\
&+&\frac{2\left[\sqrt{a(1-a)}-\sqrt{b(1-b)}\right]\left(\sqrt{a}-\sqrt{b}\right)^{2}}{\left(\sqrt{a}+\sqrt{b}\right)^{2}\left(\sqrt{1-a}+\sqrt{1-b}\right)^{4}},\nonumber
\eea

\bea
\zeta&=&\frac{16\left[\sqrt{a(1-a)}-\sqrt{b(1-b)}\right]^{4}}{\left(\sqrt{a}+\sqrt{b}\right)^{5}\left(\sqrt{1-a}+\sqrt{1-b}\right)^{5}}
+\frac{12\left(\sqrt{a}-\sqrt{b}\right)^{2}\left[\sqrt{a(1-a)}-\sqrt{b(1-b)}\right]^{2}}{\left(\sqrt{a}+\sqrt{b}\right)^{3}\left(\sqrt{1-a}+\sqrt{1-b}\right)^{5}}\nonumber\\
&-&\frac{4\left[\sqrt{a(1-a)}-\sqrt{b(1-b)}\right]^{2}}{\left(\sqrt{a}+\sqrt{b}\right)^{3}\left(\sqrt{1-a}+\sqrt{1-b}\right)^{3}}-\frac{2\left[\sqrt{a(1-a)}+\sqrt{b(1-b)}\right]
\left(\sqrt{a}-\sqrt{b}\right)^{2}}{\left(\sqrt{a}+\sqrt{b}\right)^{2}\left(\sqrt{1-a}+\sqrt{1-b}\right)^{4}},\nonumber
\eea
then we obtain
$$
\mathcal {V}=\sum_{n=1}^{5}n(d_{n})^{2}.
$$
\\
$\mathbf{Acknowledgement}$

We would like to thank the Science and Technology Development Fund of the Macau S. A. R. for generous support: FDCT 077/2012/A3, FDCT 130/2014/A3. We thank the University of Macau for generous support: MYRG 2014-00011 FST, MYRG 2014-00004 FST.


\begin{thebibliography}{}
%Type = Book
\bibitem{Andrews}
{G. E. Andrews}, {R. Askey}, {R. Roy}, {\em Special Functions}, {Cambridge University Press}, {Cambridge}, {1999}.
%Type = Book
\bibitem{Beals}
{R. Beals}, {R. Wong}, {\em Special Functions: A Graduate Text}, {Cambridge University Press}, {Cambridge}, {2010}.
%Type = Article
\bibitem{Cabanal}
{T. Cabanal-Duvillard}, {\em Fluctuations de la loi empirique de grandes matrices $al\acute{e}atoires$}, {Ann. I. H. Poincar$\mathrm{\acute{e}}$--PR} {37} ({2001}) {373--402}.
%Type = Article
\bibitem{Chen}
{Y. Chen}, {N. Lawrence}, {\em On the linear statistics of Hermitian random matrices}, {J. Phys. A: Math. Gen.} {31} ({1998}) {1141--1152}.
%Type = Article
\bibitem{Chen2012}
{Y. Chen}, {M. R. Mckay}, {\em Coulomb fluid, $Painlev\acute{e}$ transcendents, and the information theory of MIMO systems}, {IEEE Transactions On Information Theory} {58} ({2012}) {4594--4634}.
%Type = Book
\bibitem{Chihara}
{T. S. Chihara}, {\em An introduction to orthogonal polynomials}, {Dover Publications, INC.}, {New York}, {1978}.
%Type = Book
\bibitem{Gradshteyn}
{I. S. Gradshteyn}, {I. M. Ryzhik}, {\em Table of Integrals, Series, and Products: Seventh Edition}, {Elsevier(Singapore) Pte Ltd.}, {Singapore}, {2007}.
%Type = Article
\bibitem{Johansson}
{K. Johansson}, {\em On fluctuations of eigenvalues of random Hermitian matrices}, {Duke Mathematical Journal} {91} ({1998}) {151--204}.
%Type = Book
\bibitem{Lebedev}
{N. N. Lebedev}, {\em Special Functions and Their Applications}, {Dover Publications, INC.}, {New York}, {1972}.
%Type = Book
\bibitem{Mehta}
{M. L. Mehta}, {\em Random Matrices:Third Edition}, {Elsevier(Singapore) Pte Ltd.}, {Singapore}, {2006}.
%Type = Book
\bibitem{Szego}
{G. Szeg{\"o}}, {\em Orthogonal Polynomials: Fourth Edition}, {American Mathematical Society}, {Providence, RI}, {1975}.
%Type = Book
\bibitem{Wang}
{Z. X. Wang}, {D. R. Guo}, {\em Special Functions}, {World Scientific}, {Singapore}, {1989}.
\end{thebibliography}
\end{document}